\documentclass[11pt]{article}

\usepackage[a4paper,margin=1in]{geometry}
\usepackage[T1]{fontenc}
\usepackage[utf8]{inputenc}
\usepackage{lmodern}
\usepackage{amsmath,amssymb}
\usepackage{setspace}
\usepackage{hyperref}

\hypersetup{colorlinks=true,linkcolor=blue,citecolor=blue,urlcolor=blue}

\title{Gravity with a Non-Dynamical Metric in a Yang-Mills Framework}

\author{Yi Yang$^{a,b,*}$ and Wai Bong Yeung$^a$}

\date{\small
$^a$Institute of Physics, Academia Sinica, Taipei, Taiwan\\
$^b$Department of Physics, National Cheng Kung University, Tainan, Taiwan\\[0.5em]
$^*$Corresponding author: \texttt{yiyang429@as.edu.tw}}

\begin{document}
\maketitle

\begin{abstract}
General relativity assigns the spacetime metric two logically distinct roles: it defines physical
measurements and it carries the propagating degrees of freedom of gravity. We investigate
whether these roles must be assigned to the same field. We formulate a classical Yang-Mills
framework in which the metric retains its operational role in defining distances, time intervals,
causal structure, index contractions, and spacetime volume, but carries no kinetic term and no
independent propagating degrees of freedom. The metric is nevertheless varied as an auxiliary
field, producing an algebraic consistency condition rather than a propagation equation. The
gravity dynamics are instead assigned to a local GL(4,R) Yang-Mills connection with sixteen
gauge components. An explicit dictionary relates the Yang-Mills gauge potential to a
connection-like geometric variable, under which the field strength is rewritten in curvature
form. The distinctive feature of the construction is therefore not the use of GL(4,R) or a
Yang-Mills-type gravitational action by themselves, but the variational hierarchy in which a
physical measuring metric is non-propagating and algebraically constrained while the connection
carries the dynamics. We clarify the relation of this framework to Poincar\'e gauge theory,
metric-affine and Palatini formulations, pure-connection gravity, teleparallel approaches,
metric quadratic gravity, and the historical Stephenson-Kilmister-Yang/Fairchild sector. In
the restricted torsionless, metric-compatible sector, Ricci-flat geometries including the
Schwarzschild metric remain admissible exact solutions, while additional branches are retained
only as formal sectors requiring separate phenomenological assessment.
\end{abstract}

\section{Introduction}

General relativity assigns two logically distinct functions to the spacetime metric \cite{Einstein1916}. Operationally,
the metric defines physical distances, time intervals, causal structure, and spacetime volume.
Dynamically, it carries the propagating degrees of freedom of gravity, while the Levi-Civita
connection and curvature are derived from it. These two functions are normally treated as
inseparable, and most metric modifications of general relativity retain the same hierarchy even
when additional fields or higher-curvature terms are introduced
\cite{Will1993,Clifton2012,Capozziello2011,SotiriouFaraoni2010}.

The present paper asks whether this identification is necessary. Can the metric remain the
physical measuring structure of spacetime without also being the propagating gravitational
field? We explore a variational hierarchy in which the metric has no kinetic term, contains no
independent local propagating degrees of freedom, and is constrained only through an algebraic
field equation. The gravitational dynamics are instead assigned to a Yang-Mills connection \cite{YangMills1954}.
Throughout this paper, ``non-dynamical metric'' means non-propagating rather than fixed: the
metric is varied as an auxiliary variable, but its variation produces an algebraic consistency
condition instead of a differential propagation equation.

A concrete realization is obtained by taking the dynamical gauge field to be a local GL(4,R)
Yang-Mills connection $A^m{}_{n\mu}$. GL(4,R) is important because it supplies the sixteen
connection components and the transformation structure needed to relate local gauge indices to
world geometric indices. It is, however, the mechanism rather than the central physical claim.
The defining departure from metric gravity is the separation between a non-propagating metric
that retains operational meaning and a connection that carries the gravitational dynamics.

The mathematical bridge between the gauge and geometric descriptions is the explicit
relation
\[
A^m{}_{n\mu}=e^m{}_{\rho}e^{\tau}{}_{n}\Gamma^{\rho}{}_{\tau\mu}
+e^m{}_{\tau}\partial_{\mu}e^{\tau}{}_{n} .
\]
Under this dictionary, the GL(4,R) Yang-Mills field strength becomes a curvature tensor. The
construction therefore combines two complementary elements: a physical postulate concerning
the non-dynamical role of the metric, and a gauge-geometric mechanism through which the
connection dynamics acquire a gravitational interpretation.

Gauge and connection approaches to gravitation have a long history, including Lorentz and
Poincar\'e gauge theories, metric-affine gravity, Palatini formulations, pure-connection theories,
teleparallel and symmetric teleparallel gravity, and curvature-squared theories
\cite{Utiyama1956,Kibble1961,Sciama1964,Hehl1976,Hehl1995,Olmo2011,
Capovilla1989,BeltranJimenez2019,Stelle1977}. GL(4,R)-based and Yang-Mills-like gravity
constructions have also been studied previously \cite{HsuYeung1985}. We therefore do not claim novelty merely from
the choice of gauge group or from the appearance of a curvature-squared action. To our
knowledge, the specific combination developed here-a physical spacetime metric with no
kinetic term or independent propagation, varied only to impose an algebraic consistency
condition, together with a GL(4,R) Yang-Mills connection carrying the gravitational
dynamics-has not been organized in this form.

The restricted torsionless, metric-compatible sector is likewise not new as a set of vacuum
equations. It is related to the historical Stephenson-Kilmister-Yang (SKY) and Fairchild-type
quadratic-curvature system
\cite{Stephenson1958,Kilmister1961,Yang1974,Fairchild1976,Debney1978}, which is known to
admit non-unique spherically symmetric branches
\cite{Ni1975,Pavelle1976,Thompson1975,Thompson1975b}. We neither claim these restricted
equations as new nor assert that all their formal branches are physically viable. Their role here
is to provide a tractable classical sector in which the gauge-potential/connection dictionary can
be compared with familiar gravitational geometries.

The main contributions of the paper are as follows.
\begin{enumerate}
\item We separate the operational and dynamical roles of the spacetime metric by assigning it
no kinetic term and no independent propagating degrees of freedom, while retaining it as the
physical measuring structure.
\item We formulate the metric as an auxiliary, algebraically constrained field whose variation
provides a consistency condition rather than a propagation equation.
\item We assign the gravitational dynamics to a local GL(4,R) Yang-Mills connection and
construct the corresponding gauge-invariant action and field equations.
\item We establish the explicit gauge-potential/connection dictionary and rewrite the
Yang-Mills field strength in curvature form.
\item We derive the generalized energy-momentum balance law implied by this asymmetric
metric-connection variational structure and clarify the relation to both contemporary
connection-based theories and the historical SKY/Fairchild sector.
\end{enumerate}

The limitations of the present work should also be stated explicitly. We do not prove that the
full non-compact GL(4,R) gauge sector is unitary or positive definite. We do not derive a
complete weak-field phenomenology, a parametrized post-Newtonian limit, or a gravitational
wave analysis, and we do not claim uniqueness of the Schwarzschild branch. The coupling to
realistic spinor matter and the quantum consistency of the theory remain open. The present
paper is therefore a classical theoretical foundation for a broader program rather than a
completed phenomenological alternative to general relativity.

\section{The Non-Dynamical Metric as a Variational Postulate}

The defining postulate of the present framework is that the spacetime metric retains its
operational meaning but not an independent propagating dynamics. It determines physical
intervals, causal classification, index contractions, and the invariant volume element
$\sqrt{-g}\,d^4x$. It does not, however, appear with spacetime derivatives in the action and
therefore carries no kinetic term or independent local wave degrees of freedom.

The term ``non-dynamical'' should not be confused with ``fixed.'' The metric is included in the
variational principle as an auxiliary field. Functional variation with respect to
$g_{\mu\nu}$ produces an algebraic consistency condition that selects metric configurations
compatible with the Yang-Mills and matter sectors. Thus the metric is neither a prescribed
Minkowski background nor a propagating Einsteinian field. It is an algebraically constrained
measuring structure.

This separation of roles is the physical hypothesis on which the rest of the construction is
built. The metric supplies clocks, rods, causal structure, and volume, while the GL(4,R)
connection supplies the propagating gauge dynamics. In this sense, the theory does not remove
the metric from physics; it removes metric propagation from the fundamental gravitational
dynamics.

Given a metric $g_{\mu\nu}(x)$ at a point with world coordinates $x^\mu$, a set of vierbein
fields $e^a{}_{\mu}(x)$ relates the world description to a local Minkowskian frame with metric
$\eta_{ab}$. The differentials satisfy
\begin{equation}
dx^a=e^a{}_{\lambda}dx^\lambda,
\end{equation}
and hence
\begin{equation}
\eta_{ab}e^a{}_{\mu}e^b{}_{\nu}=g_{\mu\nu}.
\label{eq:vierbeinmetric}
\end{equation}
Latin indices denote local-frame components and Greek indices denote world components.
The vierbein is used here as a frame-conversion object associated with the metric; it is not
introduced as a translational gauge potential.

Because no derivatives of $g_{\mu\nu}$ occur in the action, its Euler-Lagrange equation is
algebraic in the metric once the gauge and matter fields are specified. This feature distinguishes
the present proposal from metric gravity, in which the metric equation is differential, and from
standard metric-affine theories, in which the metric is generally treated as an independent
dynamical partner of the connection.

\subsection{Physical Meaning and Testability of the Non-Dynamical Metric Postulate}

The non-dynamical metric postulate changes the physical allocation of gravitational degrees of
freedom. The metric remains the field used by matter to define physical intervals, causal
classification, and invariant volume, but it does not possess independent initial data or a wave
operator of its own. Propagation is carried instead by the Yang-Mills gauge potential. Once a
gauge-field configuration is specified, the metric is determined locally through its algebraic
consistency equation. It may therefore vary across spacetime as a response to the gauge and
matter sectors, but such variation is induced rather than independently propagating.

This distinction is in principle observable. Around an admissible background, the linearized
spectrum should be obtained by perturbing the GL(4,R) gauge potential first and then solving the
metric consistency condition for the induced metric response. The number, polarization, and
dispersion of propagating modes are therefore properties of the Yang-Mills sector rather than of
an independent metric wave equation. The same separation affects the weak-field and
post-Newtonian limits, because the metric potentials experienced by matter must be reconstructed
from the gauge-field solution rather than evolved as autonomous gravitational variables.

The postulate can be tested at several levels. A viable theory must reproduce the observed
Newtonian and Solar-System limits, yield an acceptable spectrum of gravitational-wave modes,
preserve the empirically successful universality of metric coupling to matter, and determine
whether non-Schwarzschild branches are dynamically excited or excluded by boundary and source
conditions. Cosmological solutions provide an additional test because the time dependence of the
measuring metric must follow consistently from the evolving gauge sector. These questions are
not resolved in the present foundational analysis, but they define concrete criteria by which the
proposal can be distinguished from general relativity, metric-affine gravity, and metric
quadratic gravity.

Accordingly, the non-dynamical metric assumption is not merely a change of variables or a
terminological reclassification. It is a falsifiable statement about where gravitational
propagation resides: the metric measures, while the Yang-Mills gauge potential carries the
dynamics. The remainder of this paper develops the variational and geometric structure needed
to make that separation precise.

\section{Relation to Existing Theories}

Because the present construction uses familiar words such as gauge field, connection,
curvature, vierbein, and quadratic curvature, it is important to specify its position relative to
several existing approaches. The purpose of this section is not to provide a complete review of
gauge gravity, but to make clear what is and is not being claimed.

\subsection{Poincar\'e gauge theory and Einstein-Cartan gravity}

Poincar\'e gauge theory gauges local translations and Lorentz transformations. In the usual
interpretation, the tetrad is associated with translations, the Lorentz connection is associated
with local rotations, and the resulting geometry is Riemann-Cartan geometry with curvature
and torsion \cite{Kibble1961,Sciama1964,Hehl1976,BlagojevicHehl2013,Obukhov2018}. This
framework naturally accommodates spin as a source of torsion.

The present construction is different. We do not gauge spacetime translations and do not
identify the tetrad as the translational gauge potential. The gauge field is instead
$A^m{}_{n\mu}$, a GL(4,R) Yang-Mills connection associated with changes of local geometric
frame. The metric and vierbein are used to define local frames and contractions, but they are
not the primary dynamical gauge potentials. Thus the theory is not an Einstein-Cartan or
standard Poincar\'e gauge model, even though it shares with them the broader idea that gravity
can be expressed using gauge-theoretic variables.

\subsection{Metric-affine and Palatini-type approaches}

Metric-affine gravity treats the metric and the affine connection as independent variables and
allows torsion and nonmetricity to participate in the dynamics. In the metric-affine gauge theory
of Hehl, McCrea, Mielke, and Ne'eman, the affine group and its GL(4,R) subgroup play a
central role, and the matter currents include energy-momentum and hypermomentum
\cite{Hehl1995}. More recent analyses of metric-affine dynamics emphasize that the connection
may or may not propagate depending on the invariants included in the action and on the matter
couplings \cite{Vitagliano2011}. Palatini and metric-affine $f(R)$ models similarly treat the
metric and connection independently but typically retain the metric as a dynamical field in the
variational principle \cite{Olmo2011,SotiriouLiberati2007}.

The present framework shares the refusal to impose the Levi-Civita connection from the
outset, but it differs in the role assigned to the metric. Here the metric is not an independent
dynamical partner of the connection. It is an auxiliary measuring structure that enters the
Yang-Mills action algebraically and is constrained by the metric variation. The primary
dynamical object is the GL(4,R) gauge connection. This is why we describe the construction as
connection-primary rather than metric-affine in the usual sense.

\subsection{Teleparallel and symmetric teleparallel formulations}

Teleparallel gravity and symmetric teleparallel gravity show that the gravitational dynamics of
general relativity can be represented using torsion or nonmetricity instead of curvature
\cite{NesterYo1999,BeltranJimenez2019,Bahamonde2023}. This viewpoint is often summarized
as the geometrical trinity of gravity: curvature, torsion, and nonmetricity provide different
geometrical realizations of related gravitational dynamics.

Our aim is different. We are not primarily constructing an equivalent formulation of general
relativity by choosing a curvature, torsion, or nonmetricity representation. Instead, we start
from a GL(4,R) Yang-Mills gauge field and ask how a gravitational connection and its
curvature-like representation can emerge from that gauge field. Torsionful and
Weitzenboeck-type sectors may appear in the solution space, but they are not the starting point
of the construction.

\subsection{Pure-connection formulations}

Pure-connection formulations go further than metric-affine theories by removing the metric as a
fundamental variable and reconstructing spacetime geometry from the connection or its curvature
on solutions \cite{Capovilla1989}. They share with the present work the idea that the connection
can be primary. The distinction is that our metric is not absent or merely emergent: it remains
the physical measuring structure used to define intervals, causal relations, contractions, and
volume. It is non-propagating and algebraically constrained, rather than eliminated from the
fundamental variables. The present proposal therefore lies between metric-primary and
metric-free formulations: the metric remains physically operative but is dynamically
subordinate to the Yang-Mills connection.

\subsection{Metric quadratic gravity}

Curvature-squared actions have a long history in gravitational theory and quantum gravity
\cite{Stelle1977,DonoghueMenezes2022}. In metric quadratic gravity, the metric is usually the
fundamental dynamical variable, and higher derivatives of the metric lead to additional degrees
of freedom. Such theories have been studied both as effective field theories and as candidate
high-energy completions.

The curvature-squared expression appearing below has a different origin. It is the Yang-Mills
kinetic term for the GL(4,R) gauge field rewritten through the gauge-potential/connection
dictionary. The metric does not acquire a higher-derivative kinetic term in the present
framework, because it is not treated as the primary dynamical variable. For this reason the
present model should not be classified simply as metric quadratic gravity, even though a
restricted geometrical sector resembles a quadratic-curvature system.

\subsection{Historical SKY/Fairchild sector}

The restricted torsionless, metric-compatible vacuum equations coincide formally with the
SKY/Fairchild-type system. The historical criticisms of this system are therefore relevant and
must not be ignored. In particular, the presence of non-Schwarzschild spherically symmetric
branches means that the restricted equations alone cannot be presented as a complete and
phenomenologically established replacement for general relativity.

In the present paper, this sector is used in a narrower way. It demonstrates that the GL(4,R)
Yang-Mills connection can be mapped to a familiar curvature-based equation system and that
Ricci-flat geometries, including Schwarzschild, are admissible. The additional branches are
kept as formal solutions of the restricted sector, but their physical status is not asserted. This
more modest interpretation avoids presenting the old SKY/Fairchild system as a new viable
metric theory and instead treats it as one classical sector of a broader gauge construction.
\section{Basic Physical Principles Invariant Under Local Coordinate Transformations}

On a locally flat patch around a point of our spacetime is where we do our physics. Even
though we already have a local Minkowskian system $x^a$ and $\eta_{ab}$ on that patch, we
may still have the freedom to re-label the points on that patch with different local coordinate
systems, for example, by rotating and stretching the local Minkowskian coordinate axes.

The form of the admissible local coordinate transformations depends on what physical
principles we expect to remain invariant under these coordinate changes. Here, we believe that
the law of inertia should remain intact under these expected coordinate transformations. This
means that the concept of straight lines should be preserved, as an object moving in a straight
line in one coordinate system should remain moving in a straight line in another coordinate
system. Also, light should propagate in straight lines in whatever coordinate system we are
using. Causality is also a very important concept in physics, and hence the order of points and
the ratio of segment lengths in a straight line should not change with a change in coordinate
system. And, of course, the concept of parallelness should also be preserved. Those
transformations which maintain collinearity, order of points and invariant segment ratios in
straight lines, and parallelness are the general linear transformations. General linear
transformations are sometimes grouped together as dilations, rotations, shears, and reflections.
We shall call collectively those transformations that are not rotations as strains.

\section{Marriage of the Geometric Program with the Yang-Mills Doctrine}

Here we want to emphasize that our choice of the general linear transformations as our
admissible local coordinate transformations comes from physics. It comes from our belief that
these admissible transformations should leave the above physical principles invariant. If we call
such a chosen local coordinate system a chosen local geometric setting, then we can say that
physics is assumed to be invariant under a change of local geometric setting. These
transformations form the local GL(4,R) group. It was Felix Klein who first suggested
classifying geometries by their underlying symmetry groups.

Since matter, which consists of world objects, is described by local fields with reference to a
local coordinate system, these local fields could have structures that depend on the geometric
setting chosen at that point. As we believe that the relative differences of the local fields of the
same world object at two spacetime points arising from different geometric settings are
physically meaningful only through gauge-covariant comparisons, we introduce a set of vector
bosons to counteract such variations. Similar to what was done by Yang and Mills
\cite{YangMills1954}, these vector bosons can be transformed away locally at any one point by
a suitable choice of the coordinate system at that point.

In the following, these vector boson fields are regarded as dynamical variables. Their dynamics
are fabricated so as to ensure that physics be invariant under local GL(4,R) transformations.
This is done, again by following Yang and Mills, by first constructing a Lagrangian that is
local GL(4,R) symmetric.

\subsection*{The general linear group GL(4,R)}

For a four-dimensional patch, these transformations can be carried out by $4\times 4$
invertible real matrices, either actively or passively. All these matrices form the real general
linear group of dimension 4, designated as GL(4,R). Hence GL(4,R) will be synonymous with
our symmetry group.

The GL(4,R) group has two sets of generators. The six antisymmetric generators $J_{ab}$
generate the rotations while the ten symmetric generators $T_{ab}$ generate the strains. They
satisfy the commutation relations
\begin{align}
[J_{ab},J_{cd}] &=
-i\{\eta_{ac}J_{bd}-\eta_{ad}J_{bc}-\eta_{bc}J_{ad}+\eta_{bd}J_{ac}\}, \nonumber\\
[J_{ab},T_{cd}] &=
-i\{\eta_{ac}T_{bd}+\eta_{ad}T_{bc}-\eta_{bc}T_{ad}-\eta_{bd}T_{ac}\}, \label{eq:JTcomm}\\
[T_{ab},T_{cd}] &=
 i\{\eta_{ac}J_{bd}+\eta_{ad}J_{bc}+\eta_{bc}J_{ad}+\eta_{bd}J_{ac}\}. \nonumber
\end{align}
These generators, when combined together as
\begin{equation}
M_{ab}=\frac{1}{2}(T_{ab}+J_{ab}),
\end{equation}
and with indices lowered by $\eta_{ab}$, give the compact commutation relation
\begin{equation}
[M^b{}_a,M^d{}_c]=i\delta^b_cM^d{}_a-i\delta^d_aM^b{}_c.
\label{eq:GLcomm}
\end{equation}
We introduce GL(4,R) into physics because the above elementary physical requirements point
naturally to the invariance of local geometric setting under general linear transformations.

\section{The Yang-Mills Action for Local GL(4,R)}

The Yang-Mills gauge potentials for GL(4,R) are
\begin{equation}
A_\mu=A^m{}_{n\mu}M^n{}_m .
\end{equation}
There are thus sixteen gauge bosons $A^m{}_{n\mu}$ in our theory. The Yang-Mills field
strength tensor $F_{\mu\nu}$ is
\begin{align}
F_{\mu\nu}
&=\partial_\mu A_\nu-\partial_\nu A_\mu-i[A_\mu,A_\nu] \nonumber\\
&=
\left(
\partial_\mu A^m{}_{n\nu}
-\partial_\nu A^m{}_{n\mu}
+A^m{}_{p\mu}A^p{}_{n\nu}
-A^m{}_{p\nu}A^p{}_{n\mu}
\right)M^n{}_m \nonumber\\
&\equiv F^m{}_{n\mu\nu}M^n{}_m.
\label{eq:Fdef}
\end{align}

The Yang-Mills Lagrangian, invariant under local GL(4,R) transformations, is
\begin{equation}
\mathcal{L}_{\rm YM}=\frac{1}{2}{\rm Tr}(F_{\mu\nu}F^{\mu\nu}).
\end{equation}
The calculation of the trace of the products of the generators gives
\begin{equation}
{\rm Tr}(M^b{}_aM^d{}_c)=\delta^d_a\delta^b_c .
\end{equation}
Hence the GL(4,R)-symmetric Yang-Mills action $S_{\rm YM}$ in the presence of the
non-dynamical world metric $g_{\mu\nu}$ is
\begin{equation}
S_{\rm YM}[g,A,\partial A]
=
\kappa\int \sqrt{-g}\,d^4x\,
g^{\mu\mu'}g^{\nu\nu'}
(\delta^d_a\delta^b_c)
F^a{}_{b\mu\nu}F^c{}_{d\mu'\nu'},
\label{eq:YMaction}
\end{equation}
where $\kappa$ is a dimensionless coupling constant. The total action also contains a piece
from the matter fields, making the total action
\begin{equation}
S_{\rm total}=S_{\rm YM}+\int\sqrt{-g}\,d^4x\,\mathcal{L}_{\rm matter}.
\label{eq:Totalaction}
\end{equation}

Because GL(4,R) is non-compact, the full gauge sector is not guaranteed by group theory alone
to have a positive-definite kinetic form. We therefore do not claim here to have solved all
questions of unitarity, positivity, or quantum consistency of the full theory. The present work
is restricted to the classical variational structure and to the identification of physically
interesting solution sectors.

\section{Algebraic Consistency of the Background Metric}

A particular choice of the non-dynamical world metric, and a particular set of the gauge fields and
matter fields, that together extremize the total action, will give us the physics that we are
observing in the classical world. Any world-metric configuration may be introduced at the variational level, but
only those metrics that satisfy the extremal conditions are what we are experiencing
classically. These extremal conditions are
\begin{equation}
\left.\frac{\delta S_{\rm total}}{\delta g^{\theta\tau}}\right|_A
=
\sqrt{-g}\left(
F^a{}_{c\theta\rho}F^{c\rho}{}_{a\tau}
-\frac{1}{4}g_{\theta\tau}F^a{}_{c\xi\rho}F^{c\xi\rho}{}_a
-\frac{1}{4\kappa}T_{\theta\tau}
\right)=0,
\label{eq:metriceq}
\end{equation}
\begin{equation}
\left.\frac{\delta S_{\rm total}}{\delta A^m{}_{n\nu}}\right|_g
=
D_\rho(A)\left(\sqrt{-g}\,F^{n\rho\nu}{}_m\right)
-\frac{1}{\kappa}\sqrt{-g}\,S^n{}_{m}{}^\nu=0,
\label{eq:YMeq}
\end{equation}
\begin{equation}
\frac{\delta S_{\rm total}}{\delta \hbox{matter fields}}=0.
\end{equation}
Here $D_\rho(A)$ denotes the Yang-Mills gauge covariant differentiation. The tensors
$T_{\theta\tau}$ and $S^n{}_{m}{}^\nu$ are respectively the metric energy-momentum tensor and
the GL(4,R) gauge current tensor of the source matter.

It is important to distinguish the present framework from standard Poincar\'e gauge theory. We
do not gauge spacetime translations, and we do not identify the present construction with the
Sciama-Kibble formulation. Instead, the energy-momentum tensor is defined through
variation of the matter action with respect to the independent non-dynamical metric, whereas the
GL(4,R) gauge current is defined through variation with respect to the gauge field. In this
sense, the roles of metric stress-energy sourcing and GL(4,R) gauge sourcing are kept distinct
in the present construction. Our purpose here is not to reproduce the translational sector of
Poincar\'e gauge theory, but to examine how far a Yang-Mills theory based on GL(4,R) can
account for a nontrivial classical gravitational sector when the metric is treated as
non-dynamical.

Putting the equations in words: the solved $A^m{}_{n\nu}$ from the Yang-Mills equation will
be functionals of $g_{\mu\nu}$. Plugging the solved $A^m{}_{n\nu}$ into the metric equation then
yields an algebraic consistency equation for $g_{\mu\nu}$. From this equation we select the
world metrics relevant for our classical world. The word ``select'' is used here in this
restricted algebraic sense; it does not mean that the metric carries independent propagating
degrees of freedom.

The generalized diffeomorphism identity discussed below clarifies why this algebraic metric
condition is covariantly consistent: the metric source, the GL(4,R) gauge current, and the
Yang-Mills stress tensor obey a common balance law rather than an independently closed
Einsteinian conservation law.

\section{Gauge-Potential/Connection Dictionary and Geometric Field Equations}

From the Yang-Mills fields $A^m{}_{n\nu}$ and the vierbein fields $e^a{}_\mu$, we can
construct the connection-like fields $\Gamma^\rho{}_{\tau\mu}$ by
\begin{equation}
A^m{}_{n\mu}
=
e^m{}_\rho e^\tau{}_n \Gamma^\rho{}_{\tau\mu}
+
e^m{}_\tau \partial_\mu e^\tau{}_n .
\label{eq:AtoGamma}
\end{equation}
This equation is the central dictionary between the Yang-Mills gauge language and the
geometric connection language. It is not introduced to make the metric dynamical, but to
express the GL(4,R) gauge potential in a basis adapted to a chosen local frame.

The Yang-Mills field strength tensor can then be re-expressed as
\begin{equation}
F^m{}_{n\mu\nu}=e^m{}_\lambda e^\sigma{}_n R^\lambda{}_{\sigma\mu\nu},
\label{eq:FtoR}
\end{equation}
where
\begin{equation}
R^\lambda{}_{\sigma\mu\nu}
=
\partial_\mu\Gamma^\lambda{}_{\sigma\nu}
-\partial_\nu\Gamma^\lambda{}_{\sigma\mu}
+\Gamma^\lambda{}_{\kappa\mu}\Gamma^\kappa{}_{\sigma\nu}
-\Gamma^\lambda{}_{\kappa\nu}\Gamma^\kappa{}_{\sigma\mu}.
\label{eq:Rdef}
\end{equation}
Plugging this into the Yang-Mills action gives
\begin{equation}
S_{\rm YM}[g,\Gamma]
=
\kappa\int \sqrt{-g}\,d^4x\,
g^{\mu\mu'}g^{\nu\nu'}
R^\lambda{}_{\sigma\mu\nu}
R^\sigma{}_{\lambda\mu'\nu'}.
\label{eq:curvatureaction}
\end{equation}
The Yang-Mills equation for the GL(4,R) symmetry group can thus be written as the
variation with respect to the connection-like variables $\Gamma^\rho{}_{\tau\mu}$. This
geometric representation aligns formally with quadratic-curvature theories, yet carries a
different physical meaning because the metric remains non-dynamical in the present
framework and the connection is the primary dynamical variable.

Because no kinetic term for $g_{\mu\nu}$ is introduced, the higher-derivative instability usually
associated with metric-based quadratic-curvature theories is avoided at the level of the metric
sector. This statement does not by itself prove the full positivity or quantum consistency of the
non-compact GL(4,R) gauge sector; rather, it explains why the present construction is not the
same as a conventional metric higher-derivative theory.

\section{Generalized Diffeomorphism Identity and Balance Law}

The present framework also has a useful balance-law structure. This structure is not introduced
as an independent replacement for the gauge construction above; rather, it clarifies how the
metric source, the GL(4,R) gauge current, and the gauge-field stress tensor fit together when
the metric is treated as a non-propagating auxiliary field.

The matter action is taken to depend on the non-dynamical metric, the GL(4,R) gauge field, and
the matter fields:
\begin{equation}
S_{\rm matter}=S_{\rm matter}[g_{\mu\nu},A^m{}_{n\mu},\psi].
\end{equation}
Its first variation defines the metric energy-momentum tensor and the gauge current through
\begin{equation}
\delta S_{\rm matter}
=
\frac{1}{2}\int d^4x\,\sqrt{-g}\,T^{\mu\nu}\delta g_{\mu\nu}
+
\int d^4x\,\sqrt{-g}\,S^n{}_{m}{}^\mu\delta A^m{}_{n\mu}
+
(\hbox{matter equations of motion}).
\label{eq:mattervar}
\end{equation}
Let $\xi^\mu$ be the generator of an infinitesimal diffeomorphism. Under this transformation,
the metric varies by its Lie derivative,
\begin{equation}
\delta_\xi g_{\mu\nu}=\nabla_\mu\xi_\nu+\nabla_\nu\xi_\mu,
\end{equation}
while the gauge potential, regarded as a world one-form carrying internal GL(4,R) indices,
varies as
\begin{equation}
\delta_\xi A^m{}_{n\mu}
=
\xi^\rho\nabla_\rho A^m{}_{n\mu}
+
A^m{}_{n\rho}\nabla_\mu\xi^\rho.
\end{equation}
Assuming the matter equations of motion and using the corresponding internal
gauge-current identity, diffeomorphism invariance of the matter action gives, up to sign and
normalization conventions,
\begin{equation}
\nabla_\mu T^\mu{}_\nu
=
S^n{}_{m}{}^\mu F^m{}_{n\nu\mu}.
\label{eq:Noether}
\end{equation}
Thus, in the presence of the GL(4,R) gauge interaction, the matter energy-momentum tensor
is not in general separately conserved in the naive Einsteinian sense. Its divergence is governed
by the gauge-force density carried by the field strength and the gauge current.

On the geometric side, the Yang-Mills equation of motion is
\begin{equation}
D_\rho(A)\left(\sqrt{-g}F^{n\rho\nu}{}_m\right)
=
\frac{1}{\kappa}\sqrt{-g}\,S^n{}_{m}{}^\nu,
\end{equation}
while the metric variation of the Yang-Mills action defines the geometric tensor
\begin{equation}
H_{\mu\nu}
=
F^a{}_{c\mu\rho}F^{c\rho}{}_{a\nu}
-\frac{1}{4}g_{\mu\nu}F^a{}_{c\xi\rho}F^{c\xi\rho}{}_a .
\end{equation}
Using the Yang-Mills equation together with the Yang-Mills Bianchi identity, one finds the
corresponding balance law
\begin{equation}
\nabla_\mu H^\mu{}_\nu
=
\frac{1}{2\kappa}S^n{}_{m}{}^\mu F^m{}_{n\nu\mu},
\end{equation}
again up to the normalization convention adopted in the main text. Therefore, the metric
equation
\begin{equation}
H_{\mu\nu}=\frac{1}{2\kappa}T_{\mu\nu}
\end{equation}
is covariantly consistent: the matter sector and the GL(4,R) gauge sector satisfy a common
generalized balance law, and the exchange of energy-momentum between them is governed by
the same gauge current-field-strength coupling.

\section{Classical Vacuum Sectors and Their Historical Status}

We now consider vacuum solutions of the field equations, where by ``vacuum'' we mean the
absence of matter fields except possibly at the source point. Rather than attempting to solve
the full system in complete generality, we restrict attention first to a torsionless,
metric-compatible sector. This restriction should be understood as a special classical sector of
the theory adopted for tractability, not as the most general content of the full GL(4,R)
framework.

Under this restriction, the connection $\Gamma^\theta{}_{\tau\xi}$ becomes the Levi-Civita
connection of $g_{\mu\nu}$, and the equations reduce to the Stephenson-Kilmister-Yang
equation together with the algebraic Stephenson equation:
\begin{equation}
\nabla_\tau R_{\xi\theta}-\nabla_\xi R_{\tau\theta}=0,
\label{eq:SKY}
\end{equation}
\begin{equation}
H_{\theta\tau}
=
R^\lambda{}_{\sigma\theta\rho}R^{\sigma\rho}{}_{\lambda\tau}
-\frac{1}{4}g_{\theta\tau}
R^\lambda{}_{\xi\rho\sigma}R^{\sigma\lambda\xi\rho}
=0.
\label{eq:Stephenson}
\end{equation}

Equations (\ref{eq:SKY}) and (\ref{eq:Stephenson}) are historically well known. The existence
of non-unique spherically symmetric branches in this system was historically regarded as a
serious difficulty for the SKY/Fairchild-type theories when interpreted as metric-dynamical
alternatives to general relativity \cite{Ni1975,Pavelle1976,Thompson1975,Thompson1975b}.
In the present work, we do not attempt to settle the full phenomenological status of all such
branches. Instead, our aim is to reinterpret the same restricted equations as a special sector of
a connection-primary GL(4,R) Yang-Mills framework with a non-dynamical metric.

The equations are satisfied by Ricci-flat geometries. It follows that the Schwarzschild metric is
an exact admissible vacuum solution in the torsionless, metric-compatible sector considered
here. Our claim, therefore, is not that the full theory uniquely selects Schwarzschild, but that
the framework admits Schwarzschild as a nontrivial exact vacuum metric once one
restricts to this classical sector.

It is also worth mentioning that there exists another torsionless exact solution,
\begin{equation}
ds^2
=
\left(1+\frac{G'M'}{r}\right)^{-2}dt^2
-
\left(1+\frac{G'M'}{r}\right)^{-2}dr^2
-r^2d\Omega^2 .
\label{eq:TPPN}
\end{equation}
Historically, the existence of such non-Schwarzschild branches was often regarded as a
pathology of the SKY equations. In the present framework, we retain their formal mathematical
status within the restricted equation system but postpone their physical interpretation to
companion works.

Likewise, torsionful configurations may enlarge the classical solution space. For example, a
pure-gauge configuration may be associated with a Weitzenboeck-type geometry,
\begin{equation}
ds^2=dt^2-\rho_0^2e^{2\xi t}(d\rho^2+\rho^2d\Omega^2),
\qquad
\Gamma^1{}_{01}=\Gamma^2{}_{02}=\Gamma^3{}_{03}=\xi .
\label{eq:Weitzenboeck}
\end{equation}
We mention this solution only as an illustration of the broader solution space; no observational
interpretation is asserted here.

\section{Residual Lorentz Symmetry in the Restricted Classical Sector}

A well-known difficulty for a gravity theory based on the full group GL(4,R) is the absence of
finite-dimensional spinor representations appropriate for the observed matter sector. We do not
claim to solve this problem in full generality here. Instead, we make a more limited observation
concerning the restricted classical sector considered in the previous section.

In the torsionless, metric-compatible sector, the classical solutions are described by
Levi-Civita connections compatible with the non-dynamical metric. In this situation, the effective
gauge configuration relevant to the corresponding classical metric configurations is carried by the
antisymmetric part of $A^m{}_{n\mu}$. Accordingly, the six antisymmetric generators $J_{ab}$,
which span the local Lorentz subgroup, remain relevant in this sector, whereas the symmetric
generators $T_{ab}$ do not contribute to the corresponding reduced configuration.

For this reason, the classical solution sector discussed here is effectively Lorentz-reduced. This
should be interpreted as a statement about the residual symmetry of a restricted class of
solutions, not as a complete dynamical account of spontaneous symmetry breaking in the usual
field-theoretic sense. In particular, we do not introduce a Higgs sector, an order parameter, or
a microscopic symmetry-breaking mechanism in the present work.

From this viewpoint, the restricted classical sector is compatible with the appearance of the
familiar local Lorentz symmetry relevant to observed finite-dimensional spinor fields, even
though the underlying formal construction is written in terms of GL(4,R).

\section{Scope, Physical Interpretation, and Open Problems}

The central proposal of this paper is a separation of the metric's operational and dynamical
roles. The metric remains the physical structure that defines distances, time intervals, causal
relations, contractions, and volume, but it carries no kinetic term or independent propagating
degrees of freedom. Its variation imposes an algebraic consistency condition. The propagating
gravity sector is instead supplied by the GL(4,R) Yang-Mills connection.

This assignment differs from general relativity, where the metric is the gravitational field; from
standard metric-affine gravity, where metric and connection are generally independent
dynamical variables; and from pure-connection theories, where the metric is absent as a
fundamental measuring field. The present framework keeps the metric physically operative while
making it dynamically subordinate. The gauge-potential/connection dictionary then provides
the bridge through which the Yang-Mills dynamics acquire a geometric interpretation.

In the restricted torsionless, metric-compatible sector, the framework admits Ricci-flat
geometries such as Schwarzschild. This does not imply that Schwarzschild is uniquely selected,
and the known additional SKY/Fairchild branches remain a phenomenological challenge rather
than a resolved feature. Their presence is stated explicitly so that the restricted sector is not
misrepresented as an already viable replacement for general relativity.

The present work is intended as the classical foundation of a broader Yang-Mills gravity
program. Important open issues include the consistency of the non-compact GL(4,R) gauge
sector, coupling to realistic spinor matter, the origin and interpretation of the algebraic metric
condition, weak-field and Solar-System phenomenology, gravitational-wave modes, and quantum
consistency. In particular, although gauge-field perturbations may induce effective metric
changes through the algebraic condition, the propagating mode content and observational
consequences have not yet been established.

Accordingly, the main result is not merely another GL(4,R) or curvature-squared construction.
It is the explicit realization of gravity with a physical but non-propagating metric in a
Yang-Mills framework: the metric supplies measurement and algebraic consistency, the
GL(4,R) connection supplies dynamics, the gauge-potential/connection relation supplies the
geometric bridge, and the generalized balance law supplies the associated consistency
structure. Whether this hierarchy can be developed into a complete and phenomenologically
viable theory remains to be determined.

\section*{Acknowledgments}

This work was supported by Academia Sinica and National Cheng Kung University (NCKU).
We would also like to thank Professors Friedrich W.~Hehl, James M.~Nester, and T.~C.~Yuan
for various suggestions.

\end{document}